\documentclass[pra,twocolumn,showpacs,amsmath,amssymb,superscriptaddress,longbibliography]{revtex4-2}

\usepackage{color}
\usepackage{graphicx}% Include figure files
\usepackage{dcolumn}% Align table columns on decimal point
\usepackage{bm}% bold math
\usepackage{hyperref}
\usepackage{mathrsfs}
\usepackage{soul}
\usepackage{ulem}
\usepackage{tabularx}
\usepackage{leftidx}
\usepackage[capitalize]{cleveref}
\usepackage{enumitem}
\usepackage{booktabs}
\usepackage{threeparttable} 
\setlist[itemize]{leftmargin=*}

\begin{document}

%\preprint{APS/123-QED}

\title{Simple efficient decoders for quantum key distribution over quantum repeaters with encoding}

\author{Yumang Jing}
\affiliation{School of Electronic and Electrical Engineering, University of Leeds, Leeds, LS2 9JT, U.K.}

\author{Mohsen Razavi}
\affiliation{School of Electronic and Electrical Engineering, University of Leeds, Leeds, LS2 9JT, U.K.}


\begin{abstract}
We study the implementation of quantum key distribution (QKD) systems over quantum repeater infrastructures. We particularly consider quantum repeaters with encoding and compare them with probabilistic quantum repeaters. To that end, we propose two decoder structures for encoded repeaters that not only improve system performance but also make the implementation aspects easier by removing two-qubit gates from the QKD decoder. By developing several scalable numerical and analytical techniques, we then identify the resilience of the setup to various sources of error in gates, measurement modules, and initialization of the setup. We apply our techniques to three- and five-qubit repetition codes and obtain the normalized secret key generation rate per memory per second for encoded and probabilistic quantum repeaters. We quantify the regimes of operation, where one class of repeater outperforms the other, and find that there are feasible regimes of operation where encoded repeaters---based on simple three-qubit repetition codes---could offer practical advantages.  
\end{abstract}

\date{\today}% It is always \today, today,
      
\maketitle

\section{Introduction}

Quantum repeaters (QRs) are among the key technologies that need to be developed to enable the quantum internet \cite{kimble2008quantum}. Their full implementation is, nevertheless, a daunting task, which requires progress in both experimental and theoretical fronts. An interesting avenue to pursue is to look at simple repeater structures and try to optimize their performance to offer tangible benefits. In this work, we expand on the repeater-based quantum key distribution (QKD) system studied in \cite{Jing_ErrorDet}, and propose alternative decoder structures that not only improve system performance, but also simplify its implementation. We also develop reliable approximation techniques that enable us to calculate the key rate at high nesting levels. All these will be put together to benchmark this class of quantum repeaters against alternative fully probabilistic settings \cite{duan2001,sangouard2011quantum}.

In this work, in the spirit of having an eye on near-future implementations, our focus will be on the transition from probabilistic QRs \cite{duan2001,sangouard2011quantum,razavi2009quantum} to deterministic QRs that use quantum error correction techniques only for their entanglement distillation (ED) operations \cite{jiang2009,munro2010multiplexing,zwerger2014hybrid}. In such repeaters, encoded Bell states are created along elementary links, which will then be extended to the far ends of the chain by performing entanglement swapping on encoded qubits. Such an entangled state is distilled along the way by the measurement information obtained in the entanglement swapping stage. In an earlier work \cite{Jing_ErrorDet} by our group, we show that, so long as QKD is concerned, error detection features of the code may be even more relevant than its error correction functionalities. Our results show that the majority of secret key bits come from the portion of the data that corresponds to no detected errors in either the repeater chain or decoder modules. By using error detection as a post-selection mechanism, we then improve the key rate considerably over cases when the repeater chain and decoders are treated as a black-box channel \cite{bratzik2014}. %add reference to Fig 1.

The work in \cite{Jing_ErrorDet} adopts a fully analytical approach to accurately calculate the state shared by the users and the secret key rate that can be extracted from it in the standard mode of operation. It models the error in each individual controlled not ({\sc cnot}) gate, as well as the imperfections in the initially distributed Bell states and the measurement modules. Direct calculation of the output state, even in the case of a simple repetition code, turns out, however, to be computationally challenging. In order to resolve this issue, in \cite{Jing_ErrorDet}, the authors use a linearization technique in which they work out the output state for each combination of input states at different elementary links. This reduces the computational complexity to that of a four-qubit system at its core. By accounting for the code structure, the authors then calculate the output state for different input-state combinations to obtain the final shared state between the two users. This accurate approach shows that such systems are more resilient to errors than previously thought \cite{bratzik2014}, which can make their near-future implementation more viable.

The approach in \cite{Jing_ErrorDet}, while accurate and effective for the first few nesting levels, will face computational problems at arbitrarily high nesting levels. The key reason for this is that the number of terms that need to be calculated will grow exponentially with the nesting level. At some point, as an approximation method, we need to drop the least significant terms and only keep those that majorly contribute to the key rate. In this work, we try to obtain a better understanding of such terms and devise analytical and numerical techniques that help us with reliable key rate calculations. This will then enable us to consider and analyze larger codes, e.g., five-qubit repetition codes, and compare them with simpler codes such as three-qubit repetition codes.

Another source of complexity in the analysis presented in \cite{Jing_ErrorDet} is the decoder module. The default decoder used in such settings is the one that reverses the entangling operation applied at the encoding stage \cite{bratzik2014}. The decoder module is then often composed of a number of untangling {\sc cnot} gates between different pairs of input qubits, followed by syndrome measurement and correction operations. Depending on the size of the employed code, the error analysis of the decoder part can become computationally complex. More importantly, the use of many erroneous {\sc cnot} gates has an adverse impact on the key rate as shown in \cite{Jing_ErrorDet}. This is particularly the case because the decoder modules are the last components in the setup, therefore, errors occurred in this last stage may be harder to pick up. In this work, we offer a remedy for the above problems by introducing alternative decoder structures that only rely on single-qubit measurements. This not only simplifies the QKD setup but also, by removing the major source of error from the decoding circuits, results in better performance in many practical scenarios.

%With cutting-edge quantum controlling technologies based on nitrogen-vacancy (NV) centers in diamond \cite{humphreys2018deterministic,awschalom2018quantum,kalb2017entanglement,hensen2015loophole,rozpkedek2019near,pfaff2014unconditional} and trapped ions \cite{figgatt2019parallel,wong2017demonstration,debnath2016demonstration,hucul2015modular} being developed, there is a chance that we can build QRs with, at least, simple encoding in the near future. In this situation, it will be important to have a realistic account of how such systems perform. 

Here, we first establish which decoder modules perform the best, and then, by developing several numerical and analytical methodologies, we perform a key rate analysis on QKD systems that are run over QRs with three- and five-qubit repetition codes. We account for various sources of error in the setup for each of the proposed decoders. Our main contributions are as follows:
\begin{itemize}
    \item We propose two alternative QKD decoder structures, and their corresponding post-processing, that simplify the setup and improve its performance. 
    \item We identify the terms that significantly impact the secret key generation rate, and then assess its dependence on relevant error parameters.
    \item We obtain the optimum structure for the QR setup at fixed distances, and the minimum requirements for the system to offer a positive key rate, or a rate, in bit per second per quantum memory, larger than that of a probabilistic QR.
    \item We show that, in many practical regimes of operation, the simple three-qubit repetition code is our best choice.
\end{itemize}

The paper is organized as follows. In Sec.~\ref{schematic_section}, we describe the QKD setups of interest based on the repeater protocol of Ref.~\cite{jiang2009} with four different decoder structures. By considering relevant error models for different components of the system, in Sec.~\ref{first_nesting_section}, we compare the secret key rate for different QKD decoders in the case of nesting level one for the QR setup. We then extend our results, in Sec.~\ref{higher_nesting_level_section}, to higher nesting levels by proposing several different approximation techniques. We study the dependence of the secret key generation rate on different error parameters and find the corresponding thresholds for extracting a nonzero secret key rate at different nesting levels. In Sec.~\ref{secret_key_rate_section}, we consider the entanglement generation rate of the elementary links for probabilistic and deterministic QRs, with and without multiplexing, and combine those results with the results of the previous section to obtain the total normalized secret key rates in bits per second. We also illustrate the parameter regions where one type of QR performs more efficiently than the other. Finally, we conclude the paper in Sec.~\ref{conclusion_section}.

\section{System Description}
\label{schematic_section}

\begin{figure}[htb]
\includegraphics[width=\columnwidth]{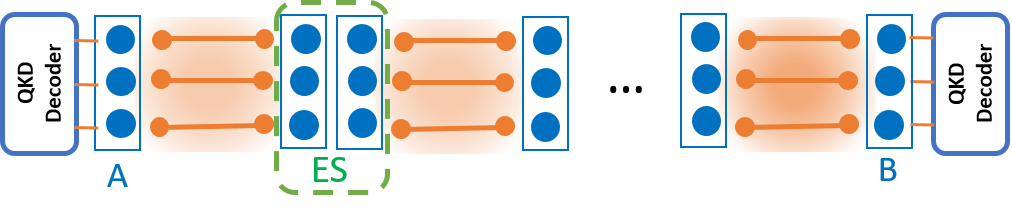}
\caption{\label{fig:setup} The schematic of the QKD setup on a repeater chain based on the three-qubit repetition code. The small oval pairs represent bipartite entangled state prepared in advance. Using remote {\sc cnot} gates, an encoded entangled state is generated across elementary links, and stored in memories represented by large ovals. The encoded entanglement is then extended across the entire link by performing entanglement swapping (ES) operations on the middle nodes. The two users will then apply decoding operation on this state to generate their raw key.}
\end{figure}

Figure \ref{fig:setup} shows the schematic of the QKD system considered in this work. Here, we use a quantum repeater with encoding \cite{jiang2009, Jing_ErrorDet} to distribute entangled states in an encoded form across the two ends of the link. We then decode such states to share a raw key between the users, Alice and Bob, from which a secret key can be extracted using postprocessing techniques. Our objective is to assess the performance of the above QKD system in the nominal mode of operation where no eavesdropper is present. In such a case, it is crucial to consider errors that stem from imperfections in the system, three major sources of which we consider in this work as follows\\

\noindent
(1) \textbf{Imperfections in initial Bell states:} The originally distributed Bell states in the QR setup are modeled as Werner states with fidelity $F_0$:
\begin{align}
\rho^{\rm W}=F_0 | \phi^{+} \rangle \langle \phi^{+} | + \frac{1-F_0}{3} (\mathbb{I}_4-| \phi^{+} \rangle \langle \phi^{+} |),
\label{original_bell}
\end{align}
where $| \phi^{+} \rangle= \frac{1}{\sqrt{2}} ( |00\rangle + |11\rangle )$, with $|0\rangle$ and $|1\rangle$ representing the standard basis for a single qubit,  and $\mathbb{I}_4$ is a $4 \times 4$ identity matrix.

\noindent
(2) \textbf{Two-qubit gate imperfections:} The {\sc cnot} gate for a control qubit $i$ and a target qubit $j$ is modeled as ~\cite{briegel1998}
\begin{align}
\label{eq:CNOTmodel}
\rho^{\text{out}} = (1-\beta) U_{i,j} \rho^{\text{in}} U_{i,j}^{\dagger} + \frac{\beta}{4} \text{Tr}_{i,j} (\rho^{\text{in}}) \otimes \mathbb{I}_{i,j},
\end{align} 
where $\rho^{\text{in}}$ ($\rho^{\text{out}}$) is the input (output) before (after) the {\sc cnot} gate, and $U_{i,j}$ represents the unitary operator corresponding to an ideal {\sc cnot} gate. The error in this two-qubit operation is modeled by a uniform depolarization of qubits $i$ and $j$, represented by identity operator $\mathbb{I}_{i,j}$, with probability $\beta$.

\noindent
(3) \textbf{Measurement imperfections:} The projective measurements to states $|0\rangle$ and $|1\rangle$ are, respectively, represented by 
\begin{align}
\label{eq:meas}
P_0 &= (1-\delta) |0\rangle \langle 0| + \delta |1\rangle \langle 1| \quad \mbox{and} \nonumber\\
P_1 &= (1-\delta) |1\rangle \langle 1| + \delta |0\rangle \langle 0|,
\end{align}
where $\delta$ is the measurement error probability. Similar measurement operators, $P_\pm$, are used for projective measurement in $|\pm \rangle = 1/\sqrt{2}(|0 \rangle \pm |1 \rangle)$ basis. 

As in \cite{Jing_ErrorDet}, here, we assume all single-qubit operations are perfect and quantum memories with infinitely long coherence times are available.

In our work, we mainly use the three-qubit repetition code as an example to illustrate our proposed techniques, where the logical qubits are encoded as
\begin{align}
    |\tilde{0} \rangle = |000\rangle \quad \text{and} \quad  |\tilde{1} \rangle = |111\rangle.
\end{align}
Some of our proposed techniques are, nevertheless, applicable to larger codes as well. Here, as an additional example, we also apply our analysis to the five-qubit repetition code, where the logical qubits are encoded as 
\begin{align}
    |\tilde{0} \rangle = |00000\rangle \quad \text{and} \quad  |\tilde{1} \rangle = |11111\rangle.
    \label{eq:5qubit}
\end{align}
This code can ideally correct up to two bit-flip errors, as compared to one in the three-qubit case. The comparison between the two codes allows us to learn how the interplay between noisy gates and stronger error-correction features affects the performance of the QKD system. 

In the following, we briefly explain the repeater model and describe different decoding modules, used in Fig.~\ref{fig:setup}, that we analyze and compare in this work.

\subsection{Quantum repeater with repetition codes}
\label{Sec:3quRep}

Here, we briefly review the protocol proposed in \cite{jiang2009} in the case of the three-qubit repetition code. The protocol with five-qubit repetition code is constructed in a similar way. For detailed description, please refer to \cite{Jing_ErrorDet}.  

The QR protocol operates in the following way. First, the codeword states, $\frac{1}{\sqrt{2}} (| \tilde{0} \rangle + | \tilde{1} \rangle)$ and $|\tilde{0}\rangle$, are locally prepared, respectively, at the left and right memory banks (large ovals in Fig.~\ref{fig:setup}), and Bell pairs are distributed between auxiliary memories (represented by small ovals in Fig.~\ref{fig:setup}) of all elementary links. Using these distributed Bell states, one can then implement remote {\sc cnot} gates, transversally, on main and auxiliary memories, after which, the encoded entangled states $\frac{1}{\sqrt{2}}(|\tilde{0}\rangle |\tilde{0}\rangle +| \tilde{1}\rangle | \tilde{1}\rangle)$ are ideally created across \textit{all} elementary links. Next, we perform entanglement swapping (ES) operations at all intermediate stations to extend the entanglement over the entire link.  This, due to the transversality of the employed code, is simply done by performing three individual Bell-state measurements (BSMs) on the corresponding pairs of physical qubits. Finally, after all ES operations, an encoded entanglement is ideally distributed between the two end users. Depending on the application in mind, the final encoded entangled state can be decoded into a bipartite state as done in \cite{bratzik2014,Jing_ErrorDet}, or be used directly as we will introduce next. In all cases, some measurement information needs to be passed to the users to identify the relevant Pauli-frame rotation on the final state.

\subsection{Decoder structures}
\begin{figure}[htb]
\includegraphics[width=\columnwidth]{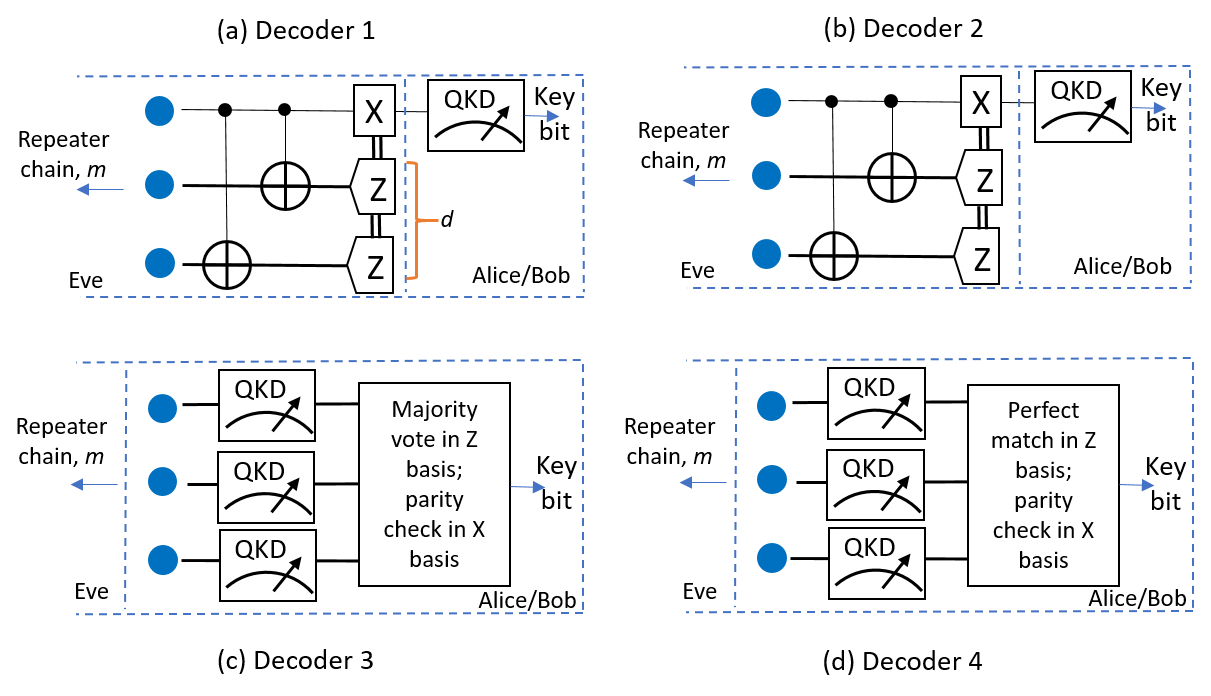}
\caption{\label{fig:dec_structure} The schematic of different decoder structures considered in this paper: (a) the original decoder proposed in \cite{Jing_ErrorDet}, where a decoding circuit is used to generate a qubit, on which a QKD measurement, either in $Z$ or $X$ basis, is performed. The decoding circuit would generate syndrome data $d$, which is used for state classification. (b) A modified version of the decoder proposed in \cite{bratzik2014}, which is very similar to (a) except that the decoder measurement outcome $d$ is not used for classification. They still use the information in $m$ in their key rate extraction. Note that, in both (a) and (b), Eve can control the decoder module, but has to pass some measurement data to users. (c) First alternative decoder, proposed in this work, where users directly measure the three qubits either all in $Z$ basis or in $X$ basis. They use majority (parity) rules, in $Z$ ($X$) basis, to decode the key bit. (d) Our second alternative decoder, which is very similar to (c), except that in the $Z$ basis a perfect match 111 (000) is mapped to bit 1 (0). In (c) and (d), we assume Alice and Bob have control over the final set of memories in their secure box.}
\end{figure}

Here, we consider four different decoder modules for the QR-QKD setup considered in this work. Figure~\ref{fig:dec_structure} shows the schematic of these decoders. In the first two decoders, Alice and Bob use the biased version of the entanglement-based BBM92 protocol \cite{bennett1992quantum,lo2005efficient} by applying QKD measurements on a bipartite state obtained via a decoding circuit. This circuit can in principle belong to a third, untrusted, party, hence Alice and Bob do not need to characterize this device in decoders 1 and 2. Its operation is based on reversing the entangling operation applied at the encoding stage. This type of decoder may also find applications in non-QKD scenarios. Figure~\ref{fig:dec_structure}(a) shows the decoder setup used in \cite{Jing_ErrorDet}, hereafter we refer to as decoder 1, in which, using {\sc cnot} gates, we first untangle the encoded entangled state, and then perform QKD measurements on the resulting bipartite state. In \cite{Jing_ErrorDet}, it is shown that the measurement information obtained at the ES, $m$, and decoding, $d$, stages can be used to separate the type of entangled states shared by the users, and consequently obtain higher key rates overall. In decoder 1, we assume that the information in $m$ and $d$ is fully used to take advantage of this classification. The second decoder, decoder 2, is shown in Fig.~\ref{fig:dec_structure}(b), where, as in decoder 1, we also apply error-correction operation to the resulting bipartite state. This decoder does not, however, pass the information obtained at the decoder stage to the user ends, and, in that sense, treats the decoder setup as a black box. Both these decoders are studied in the earlier work by our group \cite{Jing_ErrorDet}, where we show that, the particular case where no error is detected at decoding and ES stages is the major contributor to the key rate.  

In this work, we try to account for specific requirements of the QKD system to possibly come up with simpler, and as turns out more efficient, decoders. There are several observations that lead us to these alternative structures. First, we note that, so long as QKD is concerned, the purpose of the decoder module is to perform measurements in two mutually unbiased bases. Secondly, physically speaking, the three quantum memories in the two end nodes of the repeater chain are practically held in the secure boxes of Alice and Bob. The corresponding error correction/detection operations can then be performed by the two legitimate users, and not necessarily a third party. Finally, at least in the case of repetition codes, error correction/detection, or part of it, can potentially be done as part of post-processing rather than quantum mechanically.

Putting together the above points, in this work, we propose two alternative decoders and compare their performance with that of decoders 1 and 2. In both decoders, Alice and Bob, instead of manipulating their three qubits by quantum gates, directly measure them all in either $X$ or $Z$ basis. The two users choose their own basis independently, but randomly, according to the asymmetric QKD protocols \cite{lo2005efficient}. They then use classical postprocessing to assign a certain bit to their raw key bit. In decoder 3, shown in Fig.~\ref{fig:dec_structure}(c), we use the majority rule, in $Z$ basis, to replicate the error correction feature of the code against bit-flip errors. For instance, a measurement corresponding to $|101\rangle$ is mapped to bit 1. In that sense, decoder 3 can be thought of as a simplified version of decoder 2. In $X$ basis, we map the measurement outcomes that have an odd number of $|+\rangle$ states to bit 0, and all other measurement outcomes to bit 1. The former (latter) corresponds to input states that result in $|+\rangle$ ($|-\rangle$) states in the output of an ideal decoder 2. In decoder 4, shown in Fig.~\ref{fig:dec_structure}(c), we additionally apply the postselection rule proposed in \cite{Jing_ErrorDet}, where, in $Z$ basis, only measurement outcomes corresponding to no errors, i.e., $|000\rangle$ or $|111\rangle$, is kept, and all other cases are discarded. In $X$ basis, we use the same parity rule as in decoder 3. In both decoders 3 and 4, we use $m$ to postselect only cases where no error has been detected at the ES stage.

As compared to decoders 1 and 2, our alternative decoders 3 and 4 do not need to deal with the errors in the decoder {\sc cnot} gates. This certainly reduces some sources of error in the decoder, which is a sensitive part in the whole setup. Our classical postprocessing is not, however, an exact replica of that of decoder 1 as we do not use the information available in $d$ for classification. It will be interesting to see how the interplay between these two factors spans out, as we investigate in the next section.

\section{Secret key analysis for nesting level one}
\label{first_nesting_section}

In this section, we discuss the performance of the QKD system in Fig.~\ref{fig:setup} for different decoding structures of Fig.~\ref{fig:dec_structure}. As the first step, we investigate the dependence of the secret key generation rate in our QKD system on relevant error parameters in the case of one repeater node, i.e., the first nesting level. To this end, we first calculate the secret key generation rate per entangled state shared between Alice and Bob. In the asymptotic regime for an efficient entanglement-based QKD protocol, this parameter, known as the secret fraction \cite{bratzik2014}, is lower bounded by \cite{Shor-PreskillBB84}
\begin{equation}
    r_\infty{(e_b,e_p)}=\text{max}\{ 0,1-h(e_b)-h(e_p) \},
\label{eq:secfraction}
\end{equation}
where $h(p)=-p \text{log}_2 p - (1-p) \text{log}_2 (1-p)$ is the Shannon binary entropy function, and $e_b$ and $e_p$ are, respectively, the bit-flip and phase-flip error probability, or an upper bound of which, in the $Z$ basis. 

To calculate the secret fraction, in \cite{Jing_ErrorDet}, the authors develop a technique by which they analytically calculate the relevant density matrices from which the above error parameters can be calculated. Here, we use the same methodology to obtain the joint state, $\tilde{\rho}_m$, of the two memory banks held by Alice and Bob, upon observing the measurement outcome $m$ at the ES stage, and the joint state $\rho_{m,d}$, after the decoding circuit in decoder 1, upon observing, in addition to $m$, the measurement outcome $d$ at the decoding stage. The states $\tilde{\rho}_m$  and $\rho_{m,d}$ can, respectively, be obtained using Eqs. (20) and (26) in \cite{Jing_ErrorDet}, with the corresponding probability of occurrence denoted by $p_m$ and $p_{m,d}$. In our calculations, we consider a partially imperfect encoder, as modelled by Eq. (34) in \cite{Jing_ErrorDet}, where less significant cross terms are ignored. In the following, we obtain the secret fraction for each of the proposed decoders in the asymptotic regime, where an infinite number of entangled states have been shared. As mentioned before, we consider the normal mode of operation, where no eavesdropper is present, but we account for the device imperfections as modelled in Sec.~\ref{schematic_section}.

\subsection{Decoder 1}
In this decoder, the users take full advantage of $m$ and $d$ to classify their entangled states as a function of these two parameters, and extract a secret key separately from each set. In this case, the total secret fraction is given by \cite{Jing_ErrorDet}
\begin{align}
   r_{\infty}^{(1)} = \sum_{m,d} p_{m,d} r_{\infty}(e_b^{(1)},e_p^{(1)}) ,
   \label{eq:secfrac_ori}
\end{align}
where
\begin{align}
    e_b^{(1)} &= {\rm Tr}(P_0^{\rm Alice}P_1^{\rm Bob}\rho_{m,d}) + {\rm Tr}(P_1^{\rm Alice}P_0^{\rm Bob}\rho_{m,d}), \nonumber\\
    e_p^{(1)} &= {\rm Tr}(P_+^{\rm Alice}P_-^{\rm Bob}\rho_{m,d}) + {\rm Tr}(P_-^{\rm Alice}P_+^{\rm Bob}\rho_{m,d}).
\label{eq:ezex_dec}
\end{align}
The measurement operators in \cref{eq:ezex_dec} are defined according to \cref{eq:meas} with additional superscripts to specify the user. Note that, in this case, the phase-error rate is effectively the same as the bit-flip error rate in the $X$ basis.

\subsection{Decoder 2}
In the second decoder, the information in $d$ is not used for classification, but only, internally, for error correction. For each $m$, the state on which QKD measurements are performed is then given by
\begin{align}
   \rho_{m}^{(2)}= \sum_{d}{p_{d|m}\rho_{m,d}},
\end{align}
where $p_{d|m} = p_{m,d}/p_m$. The total secret fraction in this case is given by \cite{Jing_ErrorDet}
\begin{align}
   r_{\infty}^{(2)} = \sum_{m} p_{m} r_{\infty}(e_b^{(2)},e_p^{(2)} ),
   \label{eq:secfrac_bb}
\end{align}
where 
\begin{align}
    e_b^{(2)} &= {\rm Tr}(P_0^{\rm Alice}P_1^{\rm Bob}\rho_{m}^{(2)}) + {\rm Tr}(P_1^{\rm Alice}P_0^{\rm Bob}\rho_{m}^{(2)}), \nonumber\\
    e_p^{(2)} &= {\rm Tr}(P_+^{\rm Alice}P_-^{\rm Bob}\rho_{m}^{(2)}) + {\rm Tr}(P_-^{\rm Alice}P_+^{\rm Bob}\rho_{m}^{(2)}).
\label{eq:ezex_dec2}
\end{align}

\subsection{Decoder 3}
Decoder 3 uses a direct measurement on the three qubits held by each user to specify the raw key. Before calculating the corresponding error parameters, it is then important to establish the security of this structure and how we can bound the bit error rate and the phase error rate in $Z$ basis. This has been done in Appendix \ref{sec:App}, where we show that, in the ideal case, the measurement operators modelling decoder 3 are identical to that of decoder 2. We can then use a similar security proof to relate the phase error rate in the $Z$ basis to the bit error rate in the $X$ basis. In the imperfect implementation of either decoders, we end up overestimating both parameters, which is still in line with the lower bound nature of \cref{eq:secfraction}. With this in mind, the total secret fraction for decoder 3 is given by
\begin{align}
    r_{\infty}^{(3)} = \sum_{m} p_{m} r_{\infty}(e_b^{(3)},e_p^{(3)}),
   \label{eq:secfrac_altI}
\end{align}
where
\begin{align}
    e_b^{(3)} &= {\rm Tr}(\tilde{P}_0^{\rm Alice} \tilde{P}_1^{\rm Bob} \tilde{\rho}_m) + {\rm Tr}(\tilde{P}_1^{\rm Alice} \tilde{P}_0^{\rm Bob} \tilde{\rho}_m) \nonumber\\
     e_p^{(3)} &= {\rm Tr}(\tilde{P}_+^{\rm Alice} \tilde{P}_-^{\rm Bob} \tilde{\rho}_m) + {\rm Tr}(\tilde{P}_-^{\rm Alice} \tilde{P}_+^{\rm Bob} \tilde{\rho}_m)
\label{eq:exez_enc}
\end{align}
with
\begin{align}
    \tilde{P}_0 &= P_{000} + P_{100}
                + P_{010} +P_{001}, \nonumber\\
    \tilde{P}_1 &= P_{111} + P_{110}
                + P_{101} +P_{011}, \nonumber\\
    \tilde{P}_+ &= P_{+++} + P_{+--}
                + P_{-+-} +P_{--+}, \nonumber\\
    \tilde{P}_- &= P_{---} + P_{-++}
                + P_{+-+} +P_{++-}
\end{align}
being, respectively, the corresponding measurement operators to bit 0 and 1 in $Z$ basis and $X$ basis, where $P_{ijk} = P_i \otimes P_j \otimes P_k$. Note that the majority rule is used in the $Z$ basis. 

\subsection{Decoder 4}

Decoder 4 is very similar to decoder 3 with an additional post-processing step in which, in the $Z$ basis, we only accept the cases where either three 1s or three 0s have been obtained. This is inspired by the observation in \cite{Jing_ErrorDet} that the output with no error in the decoding stage is the main contributor to the key rate. In this case, the bit error rate in the $X$ basis is not necessarily an upper bound on the phase error rate for the post-selected data in the $Z$ basis. But, we can consider the worst case scenario by assuming that all the errors that we observe in the $X$ basis correspond to the post-selected part of the data in the $Z$ basis. In this case, the total secret fraction for decoder 4 is lower bounded by 
\begin{align}
    r_{\infty}^{(4)} = \sum_{m} p_{\rm succ} p_{m} r_{\infty}(e_b^{(4)},e_p^{(4)}),
   \label{eq:secfrac_altII}
\end{align}
where
\begin{align}
    p_{\rm succ}= &{\rm Tr}(P_{000}^{\rm Alice} P_{000}^{\rm Bob} \tilde{\rho}_m) + {\rm Tr}(P_{111}^{\rm Alice} P_{111}^{\rm Bob} \tilde{\rho}_m) \nonumber\\
    &+{\rm Tr}(P_{000}^{\rm Alice} P_{111}^{\rm Bob} \tilde{\rho}_m) +{\rm Tr}(P_{111}^{\rm Alice} P_{000}^{\rm Bob} \tilde{\rho}_m)
\end{align}
is the success probability for the post-selection step, i.e., detecting no error in the $Z$ basis, 
\begin{align}
    e_b^{(4)} &= \frac{ {\rm Tr}(P_{000}^{\rm Alice} P_{111}^{\rm Bob} \tilde{\rho}_m) +{\rm Tr}(P_{111}^{\rm Alice} P_{000}^{\rm Bob} \tilde{\rho}_m) }{p_{\rm succ}},
\end{align}
and
\begin{align}
    e_p^{(4)} &= \min(\frac{ e_p^{(3)} }{p_{\rm succ}},0.5).
    \label{eq:ep4}
\end{align}
The final equation gives an upper bound on the phase error rate in the $Z$ basis as explained above.

\subsection{Comparison between different decoders}

Now that we have all the ingredients to analyze all decoder settings, we can compare them in terms of their resilience to different error parameters. In \cite{Jing_ErrorDet}, the authors have already established that, by properly using the information available in $d$, decoder 1 outperforms decoder 2; please see Fig.~4 in \cite{Jing_ErrorDet}. Decoder 3, in the ideal case when there is no error in the decoder, should be identical to decoder 2, but, in the case of erroneous {\sc cnot} gates, it is expected that it outperforms decoder 2. It would not be trivial, however, if decoder 3 can outperform decoder 1 as well. Decoder 4 also, by postselecting in the $Z$ basis, can reduce the bit-flip error rate, as compared to decoder 3, but its phase-error rate bound in \cref{eq:ep4} is not necessarily tight. We have to therefore investigate if decoder 4 can ever surpass decoder 3 in terms of performance. In this section, we try to answer these questions.

\begin{figure}[t]
\includegraphics[width=\columnwidth]{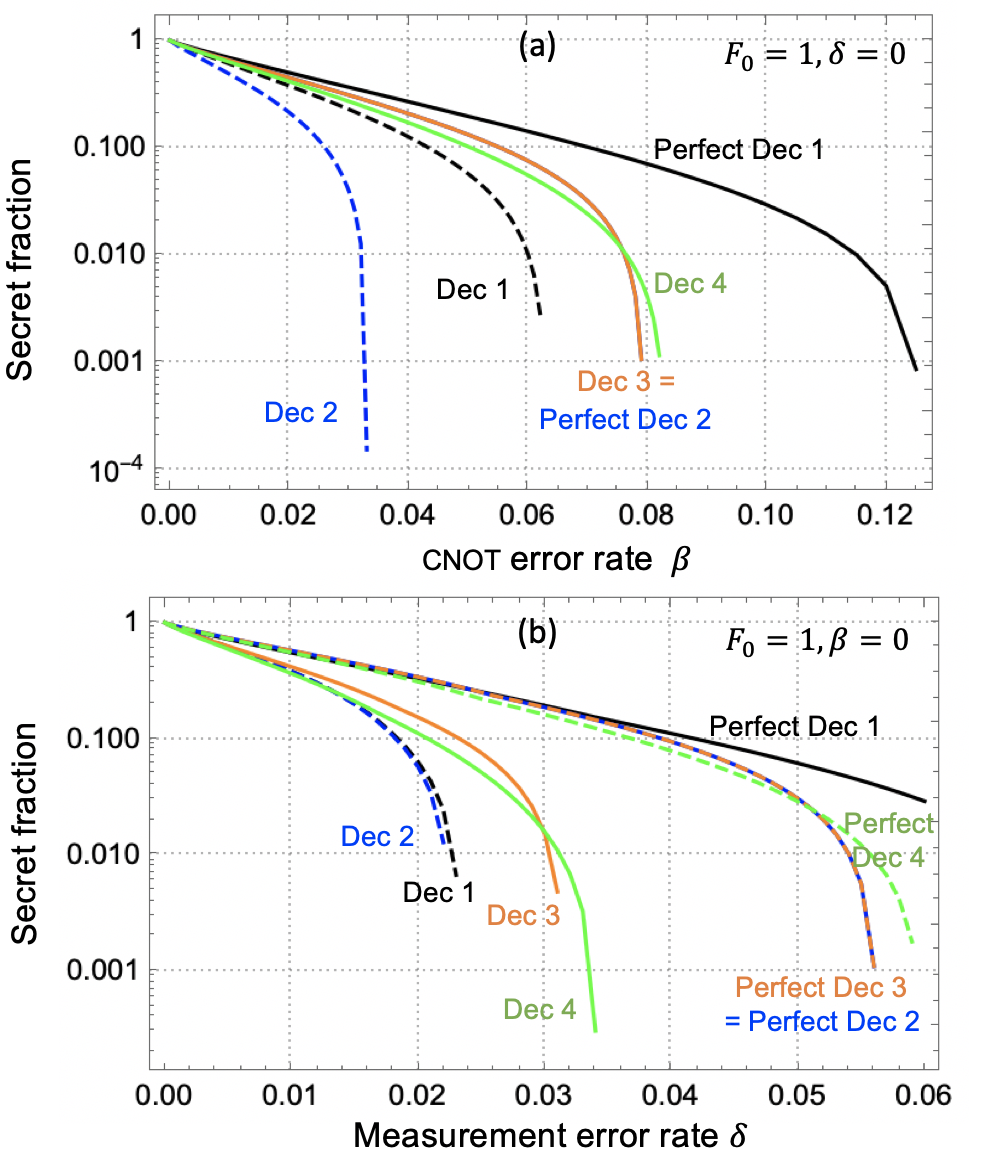}
\caption{\label{fig:alt_decoder} Secret fraction for different decoder (Dec) structures versus (a) gate error probability $\beta$ at $F_0=1$ and $\delta=0$, and (b) measurement error probability $\delta$ at $F_0=1$ and $\beta=0$. In the curves corresponding to perfect decoders, all error parameters assume their ideal values just in the decoder module; the corresponding value in the rest of the system is as the graph shows. }
\end{figure}

Figures \ref{fig:alt_decoder}(a) and (b) show the performance of different decoders, respectively, as a function of gate error probability $\beta$, at $F_0=1$ and $\delta=0$, and measurement error probability $\delta$, at $F_0=1$ and $\beta=0$. In both cases, we have also included several curves corresponding to perfect decoders as well. For instance, the perfect decoder in Fig.~\ref{fig:alt_decoder}(a) uses perfect {\sc cnot} gates as well as ideal measurement modules in its decoder circuit, whereas in the rest of the system $\beta$ can take nonzero values. We make several interesting observations from these figures, which we summarize below:

\begin{itemize}
    \item {\bf Observation 1:} Decoders 3 (orange curves) and 4 (green curves) show better performance than decoders 1 (dashed black) and 2 (dashed blue), when the imperfections in the decoder circuit are considered. In Fig.~\ref{fig:alt_decoder}(a), it is mainly the {\sc cnot} errors that make the difference. Without {\sc cnot} and measurement errors, even decoder 2 performs better than an imperfect decoder 1. This is an interesting result, which shows that the effect of {\sc cnot} errors in the decoder circuit can trump the benefits we may get from knowing the value of $d$ in decoder 1. It then follows that decoder 3 is also better than imperfect decoder 1. At $\delta =0$, this is because decoder 3 is identical to a perfect decoder 2 (see Appendix~\ref{sec:App}). But, interestingly, this also holds even for nonzero values of $\delta$ as shown in Fig.~\ref{fig:alt_decoder}(b). As a result, the maximum allowed value for $\beta$ roughly moves from 0.03-0.06, for decoders 1 and 2, to 0.08, for decoders 3 and 4. A similar behavior is seen in Fig.~\ref{fig:alt_decoder}(b), where maximum allowed value for $\delta$ roughly increases from 0.02 to 0.035.
    \item {\bf Observation 2:} We notice that, the classification versus $d$ could still play a role if all sources of error in the decoder could diminish. For instance in both figures, the curves corresponding to perfect decoder 1 offer the best performance. Also, it can be seen that, when there are no error parameters considered for decoders at all, decoders 2 and 3 perform similarly as expected by the results of Appendix~\ref{sec:App}. That said, in practice, achieving this level of perfection may not be possible, hence, so far as QKD is concerned as an application, decoders 3 and 4 are the preferred option, which not only improve the performance, but are also easier to implement.
    \item {\bf Observation 3:} In smaller error regions, decoder 3 performs slightly better than decoder 4, but eventually decoder 4, because of its postselection rule \cite{Jing_ErrorDet}, is more tolerant to errors. The good thing is that decoder 4, in terms of hardware, is exactly the same as decoder 3, and the postselection rule can be applied by software in the postprocessing steps. It is therefore feasible that, for every regime of operation, we calculate both $r_\infty^{(3)}$ and $r_\infty^{(4)}$, and pick the higher rate. In this work, the secret fraction calculated for this setup hereafter is the maximum of these two parameters denoted by $r_\infty^{\rm opt} = \max(r_\infty^{(3)},r_\infty^{(4)})$.
    \item {\bf Observation 4:} It is interesting that, in Fig. \ref{fig:alt_decoder}(b), where $\beta = 0$, decoder 3 still outperforms decoder 2 in the case of imperfect measurement modules. One may think that, given that both decoders rely on three single-qubit measurement operations, the secret fraction should be the same in both cases. Interestingly, this is not the case, and the reason for that is somehow because of the dependence of $e_b$ parameters on the location of the error as we explain next. In decoder 2, to the first order approximation, $e_b^{(2)}$ is proportional to $\delta$, corresponding to an error in the measurement on the top qubit. In decoder 3, however, we need to make at least two errors in order to have a bit flip, which means that, to the first-order approximation, $e_b^{(3)}$ is proportional to $\delta^2$. This justifies why decoder 3 outperforms decoder 2 even if $\beta = 0$. More generally, in our calculations, we realize that the position where the bit-flip occurs affects the value of $e_b^{(3)}$ in an asymmetric way. It is important then that we consider all terms in \cref{eq:exez_enc} in calculating  $e_b^{(3)}$. Note that the terms contributing to $e_p^{(3)}$ are mostly symmetric in terms of their subscripts as well as over Alice and Bob. 
\end{itemize}
\begin{figure}[t]
\includegraphics[width=\columnwidth]{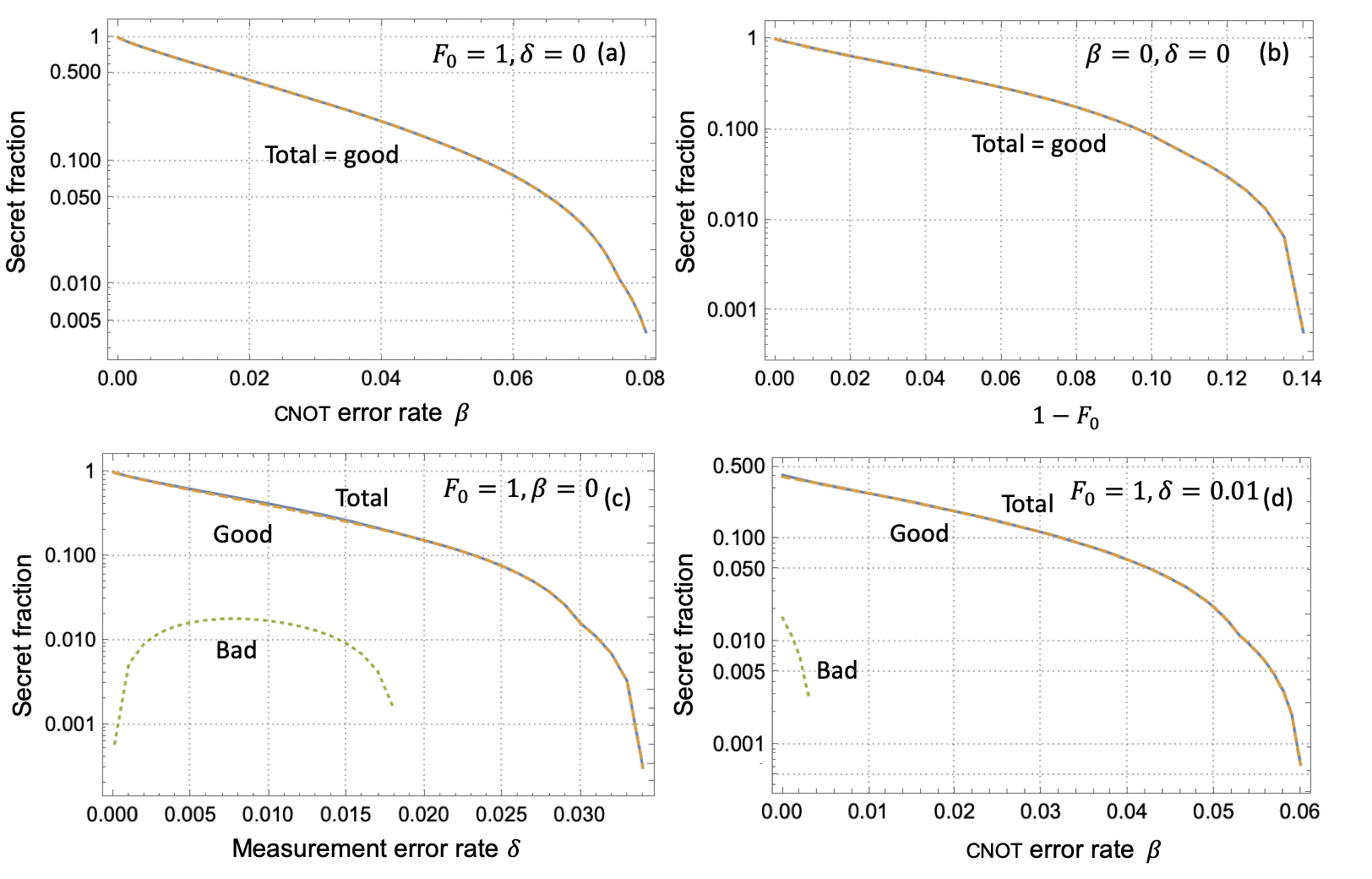}
\caption{\label{fig:goodbad} Secret fraction $r_{\infty}^{\rm opt}$ versus (a) $\beta$, at $\delta = 1-F_0 = 0$; (b) $1-F_0$, at $\delta = \beta = 0$; (c) $\delta$, at $\beta = 1-F_0 = 0$; and (d) $\beta$, at $\delta =0.01$ and $1-F_0 = 0$. The curves labelled by good correspond to the output states where no error is detected at the ES stage, whereas the bad curves are for the output states where some errors are detected at the ES stage. The curves labeled by total are the weighted sum of good and bad terms as given by \cref{eq:secfrac_altI} and \cref{eq:secfrac_altII}. }
\end{figure}

We finish this section by extending one of the key results of \cite{Jing_ErrorDet}, in using error detection as an effective postselection tool, to setups that use decoders 3 and 4. In Fig.~\ref{fig:goodbad}, we have plotted $r_\infty^{\rm opt}$ versus different error parameters, alongside the contributions from {\textit{good}} states, corresponding to no error at the ES stage, and {\textit{bad}} states, for which some error has been detected. We get very similar results to \cite{Jing_ErrorDet}, where the total secret fraction is either equal to the contribution from good states, or almost equal to it. This observation allows us in the next section to only focus on the good states, when we calculate the rate at higher nesting levels.

\section{Extension to higher-nesting levels}
\label{higher_nesting_level_section}

In order to estimate the secret key rate at higher nesting levels, we are going to use the same approach as proposed in \cite{Jing_ErrorDet}, but we modify it, using numerical and analytical approximations, so that we can manage its computational complexity. The key ingredient needed to calculate the key rate in the case of decoders 3 and 4 is the multipartite entangled state $\tilde{\rho}_m$. In this section, as explained before, we only account for the contribution from good states, and, as a representative, we only consider one particular good outcome among all that correspond to no error at the ES stage. We denote the corresponding output state to this outcome by $\tilde{\rho}_{\rm good}$. Once $\tilde{\rho}_{\rm good}$ is obtained, we can use \cref{eq:secfrac_altI} and \cref{eq:secfrac_altII} to obtain a tight lower bound on the secret fraction by ignoring the contribution from bad states. Our objective here is to get a realistic picture of what our encoded setup can achieve, and to what degree it is resilient to system errors. Exact lower bounds, which can securely be obtained in an experimental setup, are not then necessarily needed, and instead, we use tight estimates on such lower bounds to gain insight into system operation and its limitations. 

In \cite{Jing_ErrorDet}, the authors develop an analytical approach to find the joint multipartite state between Alice and Bob. In their proposed technique, they break down the initial state of the system to its core components, and apply possibly erroneous gate and measurement operations to each possible input combination separately. By using the traversality of the employed code, they then obtain $\tilde{\rho}_{\rm good}$, while avoiding the computational complexity corresponding to large multi-qubit systems. Instead, they just need to deal with a four-qubit system at a time. The number of the input terms they need to consider, however, grows exponentially with the nesting level, and practically it is very difficult to use their approach in full for nesting levels greater than three. More precisely, the entangled state between memory banks A and B, held, respectively, by Alice and Bob, for nesting level $n$, is given by (ignoring normalization factors) \cite{Jing_ErrorDet}
\begin{align}
  \tilde{\rho}_{\rm good}^{(n)} =  \sum_{\bf{j}, \bf{k}} {\bigotimes_{i=1}^3 \rho_{A_iB_i}^{\bf{j,k}}} 
\label{eq:rhoAH}
\end{align}
where ${\bf j} = [j_1,\ldots, j_{2^n}]$  and ${\bf k} = [k_1,\ldots, k_{2^n}]$ with each component taking a binary value. $\rho_{A_iB_i}^{\bf{j,k}}$ is the joint state of the $i$th memory in banks A and B if the initial state for the $2^{n+1}$ memories involved in the process is given by $\bigotimes_{l=1}^{2^n} \rho_{\rm init}^{(l)}$, where $\rho_{\rm init}^{(l)} = |j_l\rangle\langle k_l|\otimes |{0}\rangle\langle {0}|$ is the initial state of elementary link $l$. In \cite{Jing_ErrorDet}, the authors use a recursive technique to write $\rho_{A_iB_i}^{\bf{j,k}}$, at nesting level $n$, in terms of  $\rho_{A_iB_i}^{\bf{j'}}$ and $\rho_{A_iB_i}^{\bf{k'}}$, at nesting level $n-1$, with ${\bf j'} = [j_1,\ldots,j_{2^{n-1}},k_1,\ldots,k_{2^{n-1}}]$ and ${\bf k'} = [j_{2^{n-1}+1},\ldots,j_{2^{n}},k_{2^{n-1}+1},\ldots,k_{2^{n}}]$, going back to the starting point, where
\begin{align}
    \tilde{\rho}_{\rm good}^{(0)} = \frac{1}{2} \sum_{j,k=0,1}\bigotimes_{i=1}^3{\rho_{A_iB_i}^{jk}} 
\end{align}
can be calculated for each elementary link. 

As can be seen in \cref{eq:rhoAH}, the number of terms that need to be calculated for nesting level $n$ is $2^{2^{n+1}}$. This is despite the fact that  we already limit ourselves to a particular measurement outcome. For instance, at $n=3$, the number of terms is $2^{16}=65,536$, which means that our core 4-qubit calculations has to be run this many times in order to get all possible outputs. This may still sound manageable, but certainly not scalable especially if we are dealing with the analytical form of each term. 

In this work, we develop several approximation techniques to handle the computational complexity in \cref{eq:rhoAH}. By carefully analyzing each component, we find the terms that contribute negligibly to the secret fraction and can therefore be omitted. The principle behind our approximation techniques is to break the exponential growth trend and cut off the number of terms that has to be considered at each nesting level, thus improving the calculation speed dramatically. This has been achieved via analytical and numerical techniques as explained below. Using such techniques, we can also analyse larger codes in our setting, an example of which is given at the end of this section.

\subsection{Analytical approximations}
\label{sec:Analapp}
\begin{figure}[t]
\includegraphics[width=\columnwidth]{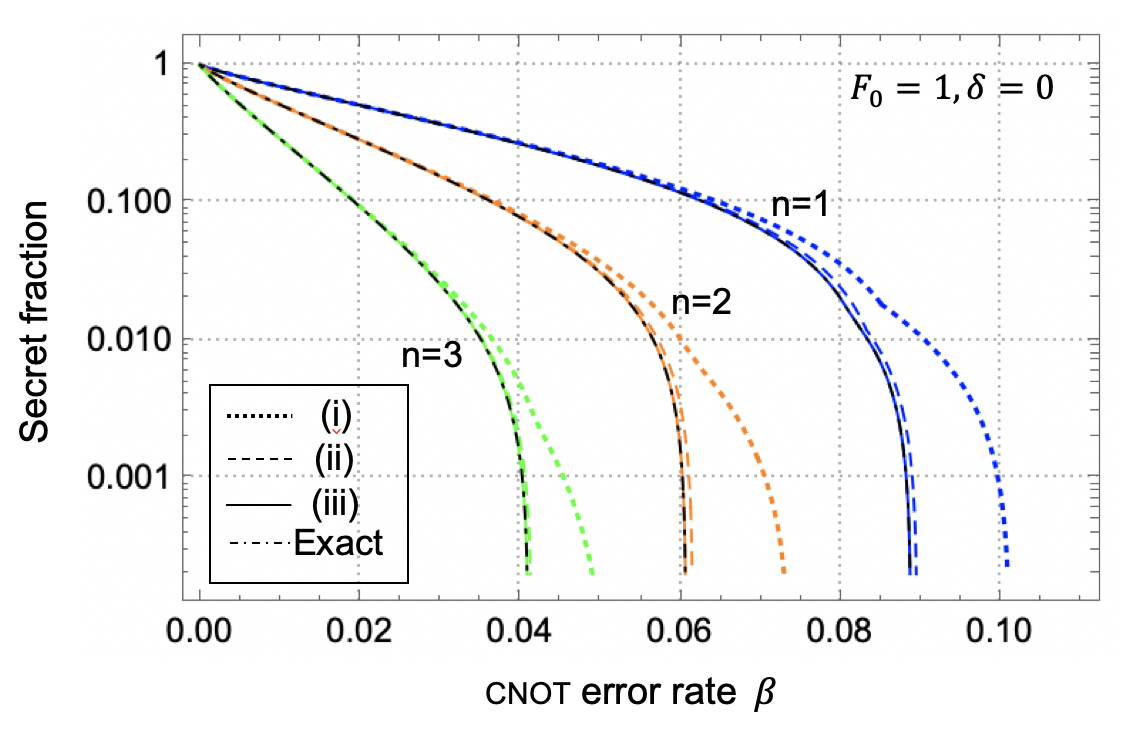}
\caption{\label{three_appro} Secret fraction  {$r_{\infty}^{\rm opt}$} versus gate error probability $\beta$ for the first three nesting levels under three different approximation methods, with initial fidelity $F_0=1$ and measurement error probability $\delta=0$.}
\end{figure}

In this section, we investigate three approximation techniques. Figure ~\ref{three_appro} gives a comparison between these three techniques and that of exact results for $n=1,2,3$ as a function of $\beta$. Our approximation method (i) is a crude one, in which, at each nesting level, we only keep four combination terms in which the initial state of all elementary links is assumed to be the same, i.e., $j_l$ ($k_l$) is the same for all values of $l$ and takes one of the possible values of 0 and 1. In other words, ${\bf j} = {\bf 0,1}$ and ${\bf k} = {\bf 0,1}$. The results, while not matching the exact curves, follows the trend very closely, at each nesting level, for small to moderate values of $\beta$. It suggests that, in this region, the contribution from the four terms with identical input states at each elementary link is the major contributor to the key rate, and all other input combinations can somehow be neglected. Approximation (i), nevertheless, cannot correctly predict the maximum value of $\beta$ at each nesting level, and only provides an upper bound on that.
Our approximation techniques (ii) and (iii), respectively, correspond to the first-order and second-order approximations of the output state $\tilde{\rho}_{\rm good}$, but with some nuances. The question is, as we deal, at lower nesting levels, with matrices corresponding to $\rho_{A_iB_i}^{\bf{j,k}}$, which of such matrices to keep at higher nesting levels, and which elements within each matrix needs to be accounted for. Note that each $\rho_{A_iB_i}^{\bf{j,k}}$ represents a two-qubit system, hence can be represented by a $4\times 4$ matrix. In method (ii), starting from nesting level one, we keep all components $\rho_{A_iB_i}^{\bf{j,k}}$ for which their matrix representation has at least one element of order $\beta$, or lower.  We also equate to zero all elements of such a matrix that are of the order of $\beta^2$ or higher. Please note that if an element has terms on the order of $\beta$ or one, that element would be fully kept. We observe strange instability in our calculations, when $\beta$ is moderately large, if we do not keep the whole element, including all higher order terms, in such cases. As a result of this purging, some combinations of $\bf{j,k}$ do not contribute to either the summation at the current level or as an input to next nesting levels. This makes the computation workload considerably lighter. Approximation method (iii) is very similar except that we keep matrices that have 
elements of order $\mathcal{O}(\beta^i)$, $i \leq 2$. In such a case, again, the full expression for the element is used even if some parts of it is of higher order than two.

As can be seen in Fig.~\ref{three_appro}, approximation methods (ii) and (iii) come very close to the exact results, with their difference to each other and the exact results becomes negligible at $n=3$. It can then be concluded that either of them would be sufficient to give us a tight estimate of the key rate at high nesting levels. This could be because, in our scheme, we only use good states for generating secret key bits. The higher order error terms can result in a larger number of errors, which would be harder to remain unnoticed, in higher nesting levels, where quite a few measurements are performed at the ES stage. This would make the contribution from terms in higher orders of $\beta$ less important. These analytical approximations give us some useful insight into which components contribute the most to the key rate. With this in mind, in the next section, we introduce a numerical approximation technique, by which we can even consider higher nesting levels.

\subsection{Numerical approximations}

For arbitrarily high nesting levels, while the analytical approximation techniques discussed previously are still applicable, the simulation speed is still severely limited by the complexity of the analytical expressions after each ES operation. Numerical techniques will then be required to find the output state in such cases. Here, based on what we learned from our analytical techniques, we propose a numerical approximation method, which is both reliable and fast. In our method, starting from nesting level one, we use the following procedure
\begin{enumerate}
    \item Calculate $\rho_{A_iB_i}^{\bf{j,k}}$ for all relevant combinations of ${\bf{j,k}}$ that we have kept in the previous nesting level (i.e., all, at $n=1$).
    \item Each $\rho_{A_iB_i}^{\bf{j,k}}$ is represented by a $4\times 4$ matrix. We collate all 16 elements for all states calculated in step 1, and sort them in the decreasing order based on their absolute values. We keep the top $N_{\rm top}$ elements, and equate the rest to zero. 
    \item We identify combinations ${\bf{j,k}}$ whose corresponding matrices have at least one nonzero element after step 2. We keep these states, and ignore the rest.
    \item Repeat the above steps for the next nesting level, until reaching the desired one.
\end{enumerate}

\begin{figure}[t]
\includegraphics[width=\columnwidth]{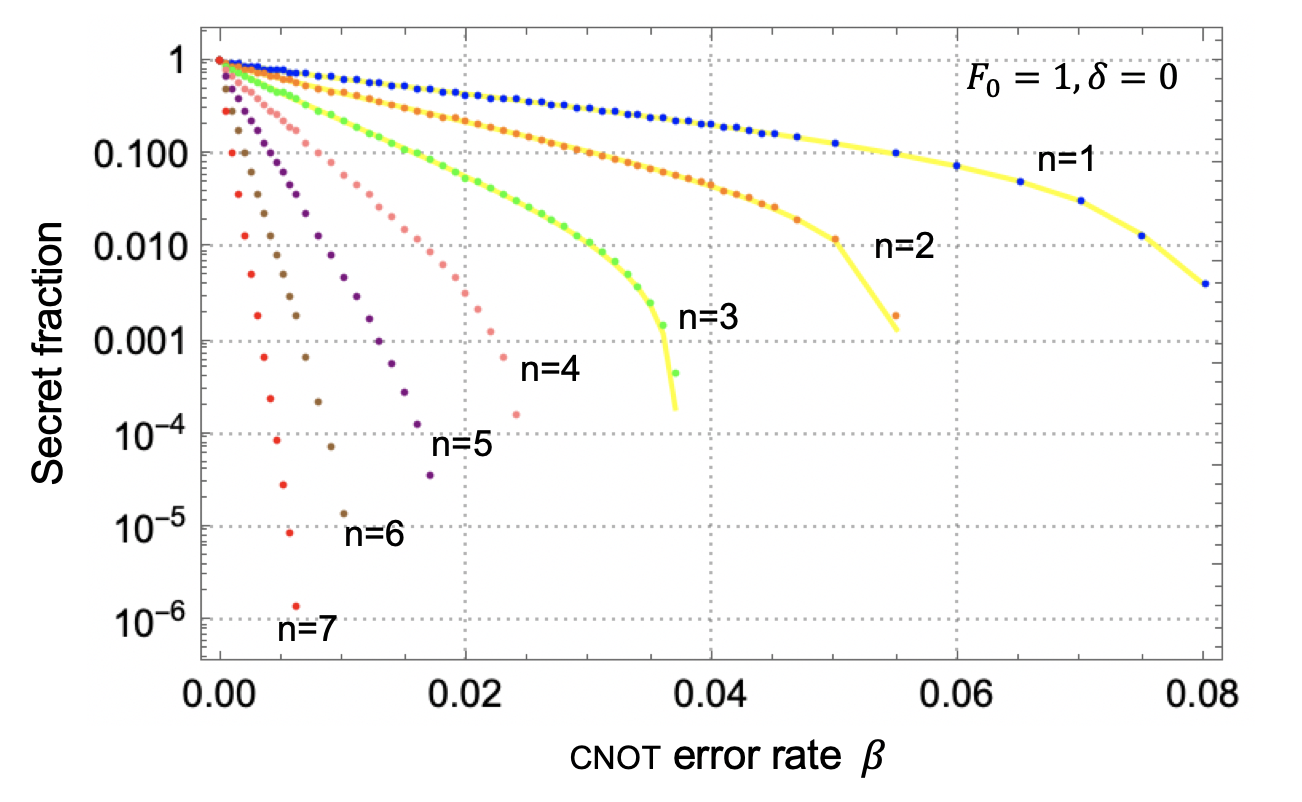}
\caption{\label{fig:higher_nesting_level} Secret fraction  {$r_{\infty}^{\rm opt}$} for three-qubit repetition code, as a function of gate error probability $\beta$ for different nesting levels, using our numerical approximation technique at $N_{\rm top} = 20$, with initial fidelity $F_0=1$ and measurement error probability $\delta=0$. Here, the errors in the encoding and decoding circuits are included. The exact simulation results for the first three nesting levels are shown (solid yellow lines) as comparison.}
\end{figure}
In Fig.~\ref{fig:higher_nesting_level}, we plot the secret fraction as a function of gate error probability $\beta$ for up to seven nesting levels, using the above algorithm at $N_{\rm top} = 20$, with initial fidelity $F_0=1$ and measurement error probability $\delta=0$. We also present the exact simulation results (solid yellow lines) for the first three nesting levels. The value of $N_{\rm top} = 20$ is chosen such that the results of our numerical approximation match the exact calculation results with high accuracy. The results of the previous section regarding the higher order terms being negligible at high nesting levels give us some assurance that the numerical results remain accurate for $n>3$ as well. We have included the results for up to seven nesting levels because, even for an elementary distance of 20~km, this already covers 10,000~km of distance for the repeater chain. This is the order of magnitude that we need for continental-scale quantum repeaters. We come back to this point in Sec.~\ref{secret_key_rate_section}. An interesting observation in Fig.~\ref{fig:higher_nesting_level} is that, even at $n=7$, the required threshold for $\beta$ is on the order of 1\%, which keeps the prospect of implementing such systems, at long distances, promising. For instance, quantum memories based on trapped ions or vacancy centers in diamond or silicon mostly meet the requirements for this setup, and can be used in early demonstrations \cite{taminiau2014universal,van2012decoherence,zhang2014protected,gaebler2016high,ballance2016high,erhard2019characterizing}. 

From Fig.~\ref{fig:higher_nesting_level}, we can obtain the maximum gate error $\beta$ that can be tolerated for extracting a non-zero secret key rate at different nesting levels. In Table II of Ref.~\cite{abruzzo2013}, a similar analysis is performed for the original quantum repeater protocol \cite{briegel1998}, also known as the BDCZ protocol after its authors.  Note that in \cite{abruzzo2013}, the authors use the gate quality $p_G=1-\beta$ as a figure of merit. Compared with their results, we notice that the quantum repeater protocol in the present work is more tolerant to gate errors at nesting levels $n \geq 3$. At low nesting levels, the BDCZ quantum repeater may, however, work better, but considering that the memory decoherence is expected to hit harder the BDCZ protocol than the encoded repeater, it is likely that the latter can perform better at lower nesting levels as well, once we consider decoherence effects. This result can be taken as an improvement of the results obtained in Ref.~\cite{bratzik2014}, where the authors conclude that the encoded QR is less tolerant against gate errors than the original QR. The change in conclusion could be mainly due to the more accurate modelling of gates and measurement modules, in our work, as well as the improvement that we get because of our classification technique, i.e., separating the cases for which no error has been detected, at the ES stage, from the rest. 

\begin{table}[h]
\caption{The simulation time for calculating the secret key fraction for different methods: Exact analytical solution, analytical approximation method (iii) in Sec.~\ref{sec:Analapp}, and the numerical approximation method at $N_{\rm top} = 20$. Here we use three-qubit repetition codes with $\beta$ as the variable, while the other two parameters are error-free. The time shown is the average time using a personal computer. The Numerical column represents the computation time per point. }\label{tab}
\begin{threeparttable}
\begin{tabular}{cccc}
  \toprule
   Nesting level & Exact $\quad$& Analytical (iii) $\quad$& Numerical \\
  \midrule
  $n=1$ & $\sim$1.5 s $\quad$& $\sim$1.5 s $\quad$& $\sim$0.06 s \\
  $n=2$ & $\sim$3.1 s $\quad$& $\sim$2.7 s $\quad$& $\sim$0.12 s \\
  $n=3$ & $\sim$65.8 s $\quad$& $\sim$9.2 s $\quad$& $\sim$0.61 s \\
  $n=4$ & $>$54852.2 s\tnote{1} $\quad$& $\sim$106.6 s $\quad$& $\sim$2.4 s \\
  $n=5$ & N/A $\quad$& N/A $\quad$& $\sim$5.3 s \\
  $n=6$ & N/A $\quad$& N/A $\quad$& $\sim$11.6 s \\
  $n=7$ & N/A $\quad$& N/A $\quad$& $\sim$31.4 s \\
  \bottomrule
\end{tabular}
\begin{tablenotes}
\footnotesize
\item[1] We stopped the simulation at this point without getting the final results.
\end{tablenotes}
\end{threeparttable}
\end{table}

Finally, it would be interesting to find out how computation time is improved using either of our analytical or numerical approximation techniques. Table~\ref{tab} shows the time consumed in each technique, including the exact analytical approach, in order to obtain the secret key fraction, at different nesting levels, in a nominal setting corresponding to Figs. \ref{three_appro} and \ref{fig:higher_nesting_level}. That is, we use {\sc cnot} error rate, $\beta$, as a variable, and fix the initial fidelity at $F_0=1$ and measurement error probability at $\delta=0$. The time shown may change if we change the parameter setting, and, in any case, they mainly represent the order of magnitude suitable for comparison, as the actual time may depend on the processor used and/or other conditions of the computing device. In our case, we have used a personal Mac machine, and we have run the simulations several times, under similar conditions, to get an average value for each point. Based on the numerical figures in Table. \ref{tab}, we notice that, as expected, the computation time scales exponentially in almost all cases, but there is a huge difference in the slope of the growth in the three cases, where for the exact analytical technique, even at $n=4$, we could not find the final answer after spending over 15 hours, whereas for the analytical approximation approach, we obtain the answer in less than two minutes. This time was only around 2~s per point in the numerical approximation case. In the end, although at low nesting levels, the computation time is about the same for all schemes, the only solution that can practically be used to assess the performance in continental-scale scenarios, or for larger codes, is the numerical one. Note that the time figures shown in the numerical case are per calculated point. The total time needed would then need to be multiplied by the number of points we are interested in. But, this additional factor would only affect the total computation time linearly. 

Now that we have sufficient tools to analyze our system, we can investigate the dependence of the key rate on another important design aspect, i.e., the employed code itself. So far, we have only dealt with the case of the three-qubit repetition code. This code is one of the simplest, and, therefore, weakest possible codes when it comes to error correction. One may wonder, if we use stronger codes, whether we get any improvement in system performance. We should bear in mind that larger codes require more gates for their encoding and decoding, and their additional error correction capabilities may be countered by the increase in the encoding errors. In the case of decoders 3 and 4, which we consider here, some key sources of error at the decoder are eliminated, but it would still be interesting to see how larger codes behave, as we investigate next.

\subsubsection{The effect of the employed code}

In this section, we find the secret fraction for a five-qubit repetition codes as described in \cref{eq:5qubit}. We will investigate if the ability of this code in correcting for up to two errors would be helpful for the QKD setup considered in this work. The setup and the protocol used is very similar to that of three-qubit code with certain obvious changes for the five-qubit case. For instance, the initial codeword state for node A, in Fig.~\ref{fig:setup}, is now ideally given by $(|\tilde 0\rangle_A + |\tilde 1\rangle_A)/\sqrt{2}$, which can be achieved by applying four {\sc cnot} gates on the state $\frac{1}{\sqrt{2}} (|0\rangle_{A_1} + |1\rangle_{A_1}) |0\rangle_{A_2} |0\rangle_{A_3} |0\rangle_{A_4} |0\rangle_{A_5}$, with $A_i$ representing the individual memories in bank A. Similar to what we have considered for the three-qubit repetition code, after accounting for errors in such gates, the codeword state for memory bank A is given by
\begin{align}
    \rho_A^{\rm in} = \rho_A^{\rm code} + \rho_A^{\rm other},  
\end{align}
where
\begin{align}
    \rho_A^{\rm code} = &\frac{1}{32}(16- 44\beta+ 49\beta^2 -25\beta^3 + 5\beta^4) \times \nonumber\\
    & (|00000\rangle_{A} \langle00000|+ |11111\rangle_{A} \langle11111|) \nonumber\\
    &+\frac{1}{2} (1-\beta)^4 (|00000\rangle_{A} \langle11111|+ |11111\rangle_{A} \langle00000|)
\label{imperfect_encoding_5}
\end{align}
contains the terms which are in the tensor product form of having the same input qubit in all rows. The state $\rho_A^{\rm other}$, which contains many more combinations of input states, is lengthy and will not be given explicitly here. Based on the observation in \cite{Jing_ErrorDet}, where it is shown that $\rho_A^{\rm code}$ part plays the major role in determining the secret fraction, here we only consider \cref{imperfect_encoding_5} and neglect the other terms. This crucially simplifies the code for further simulation.

\begin{figure}[t]
\includegraphics[width=\columnwidth]{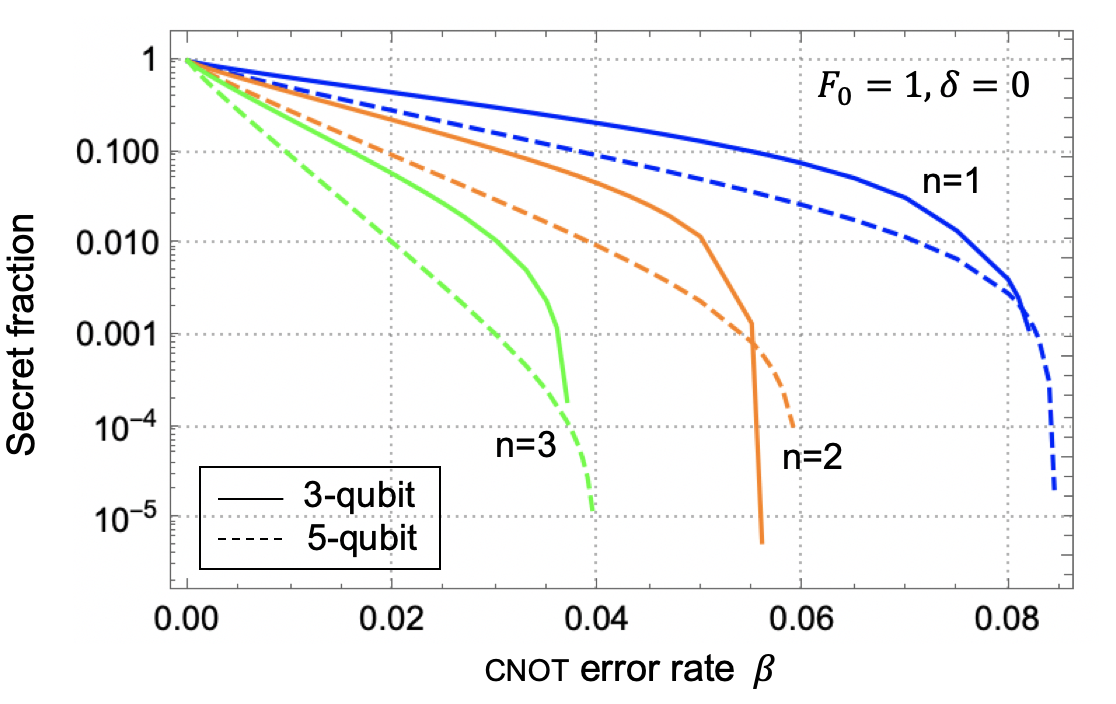}
\caption{\label{fig:5qubit} Secret fraction  {$r_{\infty}^{\rm opt}$} of QRs encoded with three-qubit repetition code (solid lines) and five-qubit repetition code (dashed lines) for the first three nesting levels as a function of gate error probability $\beta$, with initial fidelity $F_0=1$ and measurement error probability $\delta=0$.  We have used our numerical approximation method at $N_{\rm top} = 20$.}
\end{figure}

Figure~\ref{fig:5qubit} shows the secret fraction as a function of gate error $\beta$ for QRs with five-(dashed lines) and three-qubit repetition codes (solid lines), at $F_0=1$ and $\delta=0$, up to $n=3$. We notice that, initially, the protocol with three-qubit repetition code generates more keys. This is expected as, at low values of $\beta$, there are not that many errors and the three-qubit code can detect them similarly to the five qubit code, without imposing additional encoding errors. However, with the increase in $\beta$, the protocol with five-qubit repetition codes begins to show advantage over the three-qubit code since it can tolerate more errors. We have to wait and see if this possible advantage at higher error rates is of any practical relevance. We give an answer to this question in the following section. 

\section{Secret key rate for the repeater chain}
\label{secret_key_rate_section}

The most practical figure of merit for a QKD system is often represented by its total secret key generation rate, $R$, in bits per unit of time. Thus far, we have only focused on the secret fraction, which gives us the probability of generating a secret key once a multipartite entangled state is shared with the users. In order to obtain the total secret key generation rate, we need to multiply the secret fraction by the entanglement generation rate $\gamma$. In this section, we account for the latter factor, in two possible implementations of the setup in Fig.~\ref{fig:setup}, as well as the corresponding fully probabilistic quantum repeater setups. This allows us then to specify the regions in which each setup could offer a better performance.

In all cases considered in this section, we assume that a DLCZ-like protocol \cite{duan2001} is used to distribute entanglement over the elementary links. In this scheme, entangled memory-photon pairs are generated simultaneously at each elementary node, the photons are coupled into optical fibers and interfere in the middle of each segment. A successful BSM, which is classically communicated to the two end nodes of the elementary link, projects the corresponding memory qubits into an entangled states. In our work, we assume that the success probability for each entangling attempt is given by \cite{sangouard2009,sangouard2011quantum}
\begin{align}
 P_0=\frac{1}{2} p^2 \eta_{\rm ch}^2 \eta_d^2,   
\end{align}
where $p$ represents the probability of generating the initial memory-photon entanglement and the coupling efficiency of a photon into  the optical fiber, $\eta_d$ accounts for the detector efficiency and its corresponding coupling efficiency, and $\eta_{\rm ch}=\exp [-\frac{L_0}{2L_{att}}]$ is the transmitivity of a photon through half of the elementary link with length $L_0$. $L_{\rm att}$ is the attenuation length of the channel, where for standard optical fibers is around 22~km.  Also, ignoring the measurement time and the interaction time between memories and photons, each entangling attempt as above would take
\begin{align}
   T_0=L_0/c, 
\end{align}
which includes the initial transmission of the photon and the classical communication to verify the success, with $c=2\times 10^5$ km/s being the speed of light in fiber.  With the successful probability for generating one Bell pair being $P_0$, the average waiting time for generating $N$ Bell pairs is given by 
\begin{align}
    <T>_N = T_0 Z_N(P_0)
\end{align}
where $Z_N(P_0)$ is the average number of trials required to distribute $N$ Bell pairs given by \cite{bernardes2011}
\begin{align}
    Z_N(P_0)&=\sum_{k=1}^N\left( \begin{array}{c} N\\k \end{array} \right) \frac{(-1)^{k+1}}{1-(1-P_0)^k}. 
\end{align}

Based on the above entanglement distribution protocol, we now consider several QR protocols based on error correction, with and without multiplexing, and probabilistic ES operations, and compare them together.

\subsection{Encoded QR with no multiplexing}
Here, we consider, the setup in Fig.~\ref{fig:setup}, with minimal number of logical quantum memories, that is, $2q$ per memory bank for a $q$-qubit repetition code with $q=3,5$. The factor two accounts for the memories used for initial entanglement distribution (small ovals) and those used for the remaining steps (large ovals). The total number of memories, at nesting level $n$, in the setup is then given by $2^{n+2}q$. In this setting, whenever, at any intermediate node, the initial entanglement over its adjacent links are prepared, we can go ahead and perform the corresponding ES operation at that node. By this technique, on average it will take $Z_{N_{\rm det}}(P_0)$, for ${N_{\rm det}}= q \times 2^n$, until we have all elementary links entangled, and done the corresponding ES operations. The entanglement generation rate, i.e., the number of encoded entangled states shared per second is then given by 
\begin{align}
    \gamma_{\text{det}}=\frac{1}{<T>_{N_{\rm det}}}.
    \label{ent_rate}
\end{align}
The subscript $\text{det}$ refers to the deterministic QR considered in the present work. The normalized secret key generation rate is then given by
\begin{align}
    R_{\text{det}}=\frac{ r_{\infty}^{\rm opt} \times \gamma_{\text{det}} }{ 2^{n+2}q},
    \label{eq:RateDet}
\end{align}
where $r_{\infty}^{\rm opt}= \max(r_{\infty}^{(3)},r_{\infty}^{(4)})$. 

Note that, in practice, each physical memory module may contain multiple logical qubits. For instance, for nitrogen vacancy centers in diamond, both electronic and nuclear spins can be used as a qubit. In such cases, the normalized rate in \cref{eq:RateDet} can be modified to account for this factor. 

\subsection{Encoded QR with multiplexing}
With the probability for successfully generating an entangled pair $P_0$ being small, a large number of attempts will be needed before one elementary link is ready for use. One could, however, do this entangling process in parallel across many pairs of memories. This multiplexing operation improves the rate and resilience to decoherence \cite{razavi2009quantum,QRMux-Kuzmich} at the price of requiring significantly more physical resources to minimize the required temporal resources. Here, we consider the case where there are a large number of memories $N_m$ per station satisfying $N_m P_0 \gg 1$. Using this multiplexing technique, we can ensure that, after every attempted cycle of duration $T_0$, there are enough entangled pairs generated to immediately perform the QR swapping and QKD measurement operations. The entanglement generation rate, for this continuously running system, is given by
\begin{align}
    \gamma_{\text{det}}^{\text{Mux}}= \frac{N_m P_0}{q T_0},
\end{align}
which leads to the normalized secret key generation rate as follows
\begin{align}
    R_{\text{det}}^{\rm Mux}=\frac{ r_{\infty}^{\rm opt} \times \gamma_{\text{det}}^{\text{Mux}} }{4N_m \times 2^{n}} = \frac{ r_{\infty}^{\rm opt} P_0}{ 2^{n+2} q T_0}.
\end{align}
\begin{figure}[t]
\includegraphics[width=\columnwidth]{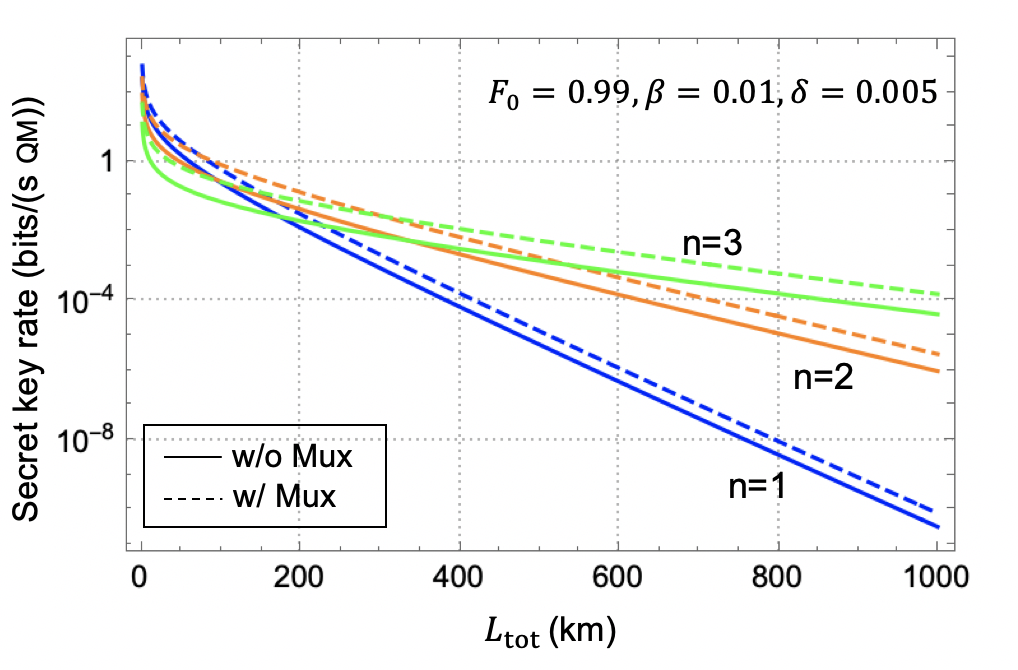}
\caption{\label{fig:multiplexing} Normalized secret key rates for the encoded QRs with/without multiplexing for the first three nesting levels as a function of the total distance, with initial fidelity $F_0=0.99$, gate error probability $\beta=0.01$ and measurement error probability $\delta=0.005$. The secret key rate is calculated for the better of decoders 3 and 4 at $p=0.5$, $\eta_d = 0.9$, and $L_{\rm att} =$22~km using our numerical approximation technique at $N_{\rm top} = 20$.}
\end{figure}
In Fig. \ref{fig:multiplexing}, we plot the normalized secret key rate for QRs with 3-qubit repetition code with and without multiplexing as a function of the total distance $L_{\rm tot} = 2^n L_0$. We assume an initial fidelity $F_0=0.99$, gate error probability $\beta=0.01$, and measurement error probability $\delta=0.005$. As for other parameters, we have assumed $p=0.5$, which is achievable for cavity-enhanced memories \cite{riedel2017deterministic}, and $\eta_d = 0.9$ \cite{marsili2013detecting}. For the chosen parameters, we can generate  non-zero key rates up to the third nesting level. Note that according to Fig.~\ref{fig:multiplexing}, the optimum distance for elementary links is about 50~km. Since the memory coherence time and dark count rate of detectors are not taken into account in this analysis, we do not see the typical cut-off security distance beyond which secure key exchanges is not possible. It is expected that a coherence time on the order of $10 T_0$ and $10 <T>_{N_{\rm det}}$ are, respectively, needed for the proper operation of the system with and without multiplexing \cite{piparo2013long}. We notice that, as expected, the multiplexing helps increase the secret key rate. The higher the nesting level, the more visible this increase is. But, even with multiplexing, the total rate achievable by the system is rather low. For instance, at a total distance of 800~km, we would need around 1000 quantum memories to obtain a total key rate on the order of bits per second. This is comparable with what one may achieve with probabilistic quantum repeaters. Next, we will consider this class of quantum repeaters for a more quantitative analysis. 

\subsection{Probabilistic quantum repeaters}
The most feasible implementations of quantum repeaters rely on probabilistic operations for the initial entanglement distribution as well as further ES operation \cite{yu2020entanglement}. Initially proposed by Duan, Lukin, Cirac and Zoller (DLCZ) \cite{duan2001}, it soon found various alternatives \cite{sangouard2011quantum, Amirloo-Razavi}. Here, we use a generic model for this class of quantum repeaters to enable a fair comparison with encoded systems when it comes to QKD as an application. One key difference is in the fact that the implementation of probabilistic ES is not based on gate operations. Instead, a probabilistic photonic ES can be achieved by converting back the state of quantum memories to single photons and then do BSMs on the corresponding photons. We can therefore neglect all gate and measurement errors in probabilistic repeaters, and only consider imperfections in the initially distributed Bell states. The main drawback of such a protocol is that probabilistic BSMs increase the waiting time and reduce the rate. The resilience to decoherence would also be lower, requiring coherence time on the order of $10 \times L_{\rm tot}/c$, because of additional transmission delays, even if multiplexing is used \cite{piparo2013long}. 

\begin{figure}[htbp]
\includegraphics[width=.8\columnwidth]{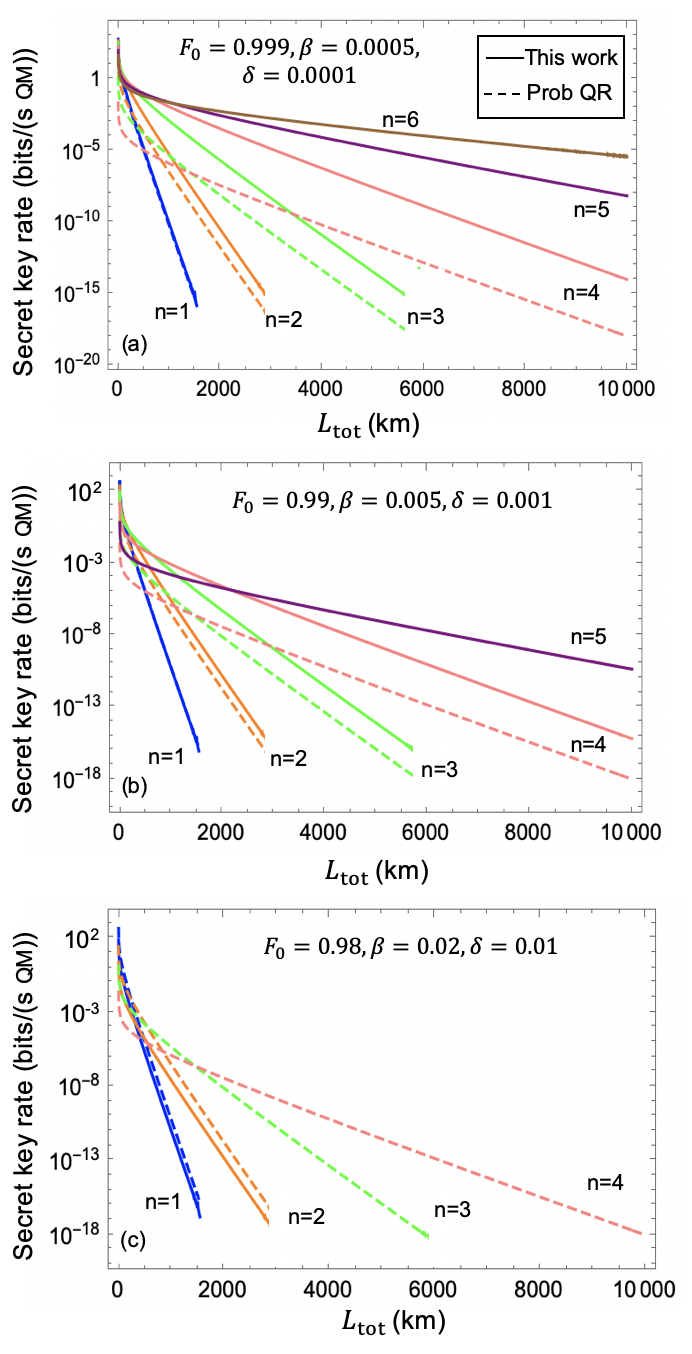}
\caption{\label{fig:probabilistic} Normalized secret key rates for QRs with encoding (solid, three-qubit code) and probabilistic QRs (dashed) in the absence of multiplexing for up to six nesting levels as a function of the total distance, with different error parameters: (a) $F_0=0.999$, $\beta=0.0005$ and $\delta=0.0001$; (b) $F_0=0.99$, $\beta=0.005$ and $\delta=0.001$; (c)$F_0=0.98$, $\beta=0.02$ and $\delta=0.01$. Other parameters are as in Fig.~\ref{fig:multiplexing}. In the encoded repeater case, the secret key rate is calculated for the better of decoders 3 and 4 using our numerical approximation method at $N_{\rm top} = 20$.}
\end{figure}

Based on above assumptions, in this work, the ES operation is modelled as follows. If, by nesting level $n$, the entangled states on $ab$ and $cd$ links is diagonal in the Bell basis and is given by $ \rho_{ab}=\rho_{cd}= A_n |\phi^{+} \rangle \langle \phi^{+}| + B_n |\phi^{-} \rangle \langle \phi^{-}| + C_n |\psi^{+} \rangle \langle \psi^{+}| + D_n |\psi^{-} \rangle \langle \psi^{-}|$, the resulting state between $a$ and $d$ after a BSM on $b$ and $c$ can still be written in the Bell diagonal form with the following new coefficients \cite{abruzzo2013}
\begin{align}
    A_{n+1} &= (A_n^2+B_n^2+C_n^2+D_n^2), \nonumber\\
    B_{n+1} &=2(A_nB_n+C_nD_n), \nonumber\\
    C_{n+1} &=2(A_nC_n+B_nD_n),\nonumber\\
    D_{n+1} &=2(A_nD_n+B_nC_n).
\end{align}
The initial state of the elementary links in our analysis is given by \cref{original_bell}. The successful probability for the ES operation is assumed to be 
\begin{align}
    P_{\text{ES}} = \frac{1}{2} p_m^2 \eta_d^2,
\end{align}
where $p_m$ is the reading and coupling efficiency of memories, which, for simplicity, here we assume $p_m = p$. The entanglement generation rate for such a protocol can be derived as \cite{sangouard2011quantum}
\begin{align}
    \gamma_{\text{prob}}& = \frac{1}{ <T>_{N_{\rm prob}}} P_{\text{ES}}^n 
\end{align}
for $N_{\rm prob} = 2^n$. Here, we assume the ES success probability is the same for all nesting levels. The normalized secret key rate per memory is then given by
\begin{align}
  R_{\text{prob}}=\frac{ r_{\infty}^{\rm prob} P_{\text{click}} \gamma_{\text{prob}}}{2^{n+1}},  
\end{align}
where $P_{\text{click}}=\eta_d^2$ is the success probability for performing QKD measurements (twofold coincidence), and 
\begin{equation}
    r_{\infty}^{\rm prob} = r_\infty(C_n+D_n, B_n + D_n)
\end{equation}
at nesting level $n$.

One can similarly find out the corresponding key rate equations in the case of multiplexed repeaters; see, for instance, \cite{razavi2009quantum}. The results are very similar to that of Fig.~\ref{fig:multiplexing}. For the sake of comparison that we are pursuing in this paper, we obtain similar results if we use encoded and probabilistic QRs both with, or without, multiplexing. Given that, for early demonstrations of quantum repeaters, quantum memories are quite precious, next we compare the two systems only in the case of no multiplexing for which fewer memories are needed.

\subsection{Optimal QRs in different parameter regions}

Figure \ref{fig:probabilistic} shows the secret key rate for encoded and probabilistic QRs for three sets of parameters. These three sets represent different degrees of reliability for our quantum gates and measurements. This mainly affects the encoded repeater case, and we notice that, in all three cases, the probabilistic QR (dashed lines) can only offer a key up to nesting level four, while the encoded QR (solid lines) can offer better rate-versus-distance scaling by using higher nesting levels. In Figs.~\ref{fig:probabilistic}(a) and (b), corresponding to low-error and moderately-low-error regimes, we notice that the QR with encoding offers higher key generation rates than the probabilistic one. This advantage increases with the nesting level to the point that, at $n=4$, the QR with encoding improves the key rate by more than three orders of magnitude. In lower error regime with $F_0=0.999, \beta=0.0005, \delta=0.0001$, the encoded QR can generate secret keys up to $n=6$ corresponding to 64 elementary links. If we multiply the number of QMs required at this nesting level by the normalized key rate, the total key rate is $\sim 10^{-2}$ bits per second at 10,000~km, which is comparable to what currently most advanced fiber-based QKD techniques can achieve at distances below 1000~km \cite{TFQKD509km, Curras2021_npjQI}. If the error parameters are increased by one order of magnitude, as in Fig.~\ref{fig:probabilistic}(b), secret key can only be extracted up to $n=5$ for encoded QRs, and the generated key will be reduced by two orders of magnitude as compared to Fig.~\ref{fig:probabilistic}(a) at $n=5$. In the higher error regime, shown in Fig.~\ref{fig:probabilistic}(c), the probabilistic QR shows better performance in most distances. This is mainly because of the errors involved in the gates, which makes the use of error correction codes less effective. In Fig.~\ref{fig:probabilistic}(c), the QR with three-qubit encoding can only generate secret keys up to $n=2$ with the specified error parameters. We conclude that the QR with encoding will only be useful when the error rates are moderately low.

\begin{figure}[htbp]
\includegraphics[width=.75\columnwidth]{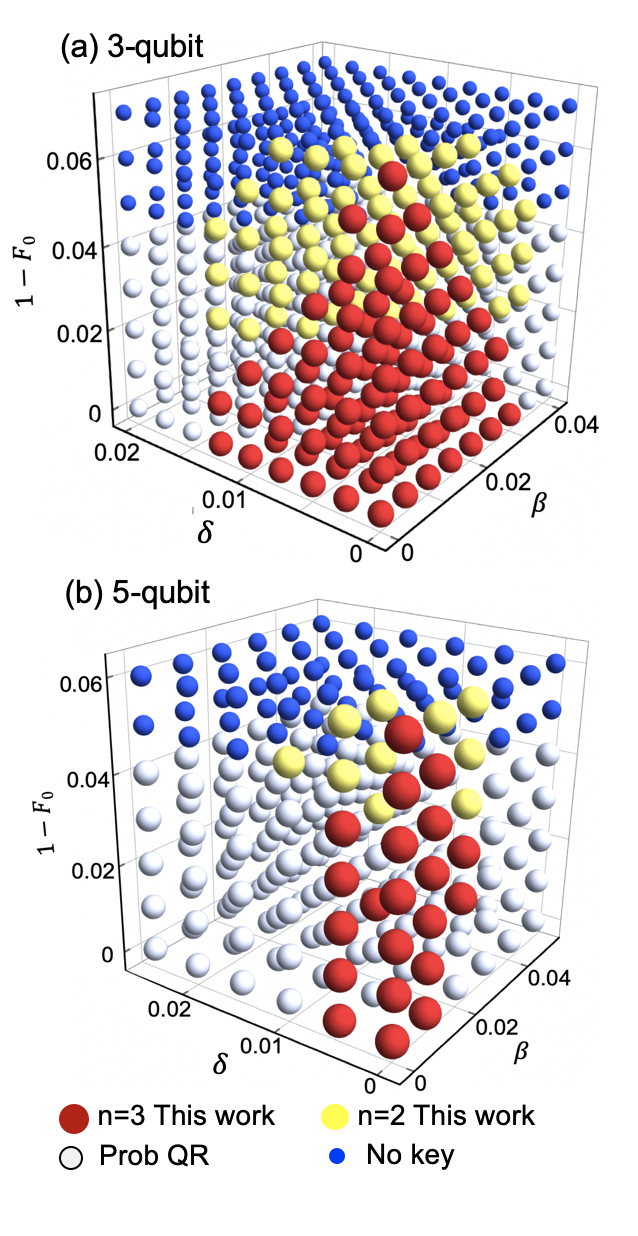}
\caption{\label{fig:3dplot}  The region plots showing the distribution of the optimal QR protocol in a three-dimensional parameter space at $L_{\rm tot}=1000{\rm km}$ for (a) three-qubit repetition code and (b) five-qubit repetition code. Other parameters are as in Fig.~\ref{fig:multiplexing}. In the encoded repeater case, the secret key rate is calculated for the better of decoders 3 and 4 using our numerical approximation method at $N_{\rm top} = 20$.}
\end{figure}

The examples in Fig.~\ref{fig:probabilistic} imply the existence of operation regions in which one or the other QR could offer a better performance.  In Figs.~\ref{fig:3dplot}(a) and (b), we have, respectively, specified these regions, at a fixed distance of 1000~km, for three-qubit and five-qubit repetition codes. The choice of 1000~km corresponds to possibly near-term implemetations of QR systems that outperform no-repeater systems. In both figures, we have highlighted which QR structure offers the higher key rate, if any, as a function of our three error parameters $1-F_0$, $\beta$, and $\delta$.  We identify four regions in Figs.~\ref{fig:3dplot}(a) and (b): 
\begin{itemize}
    \item Region 1: For low gate and measurement error probabilities, when initial fidelity of Bell states is high, the third nesting level of QRs with encoding dominates.
    \item Region 2: For the same region of gate and measurement error probabilities, when initial fidelity becomes worse, the second nesting level of QRs with encoding is more favourable.
    \item Region 3: For slightly higher gate and measurement error probabilities, the encoded QR loses its advantage, and probabilistic QRs are the best option.
    \item Region 4: For high error probabilities and low initial fidelity of the original entangled states, it is not possible to generate a secure key with either of QR protocols.
\end{itemize}

It is interesting to note that the region that the three-qubit code outperforms the probabilistic QR is larger than that of the five-qubit code. In fact, by the time that, according to Fig.~\ref{fig:5qubit}, the five-qubit QR outperforms the three-qubit one, both encoded QR structures perform worse than the probabilistic repeater. This would suggest then it is likely that the best error correction codes for QKD purposes are the simplest ones, and it may not be necessary to overcomplicate the system by employing larger codes. In fact, so long as QKD is concerned, the practical choices seem to be between a probabilistic structure for the quantum repeater versus the three-qubit code in the encoded structure. We mention that we do not consider all sources of imperfection in our analysis, and it may be ill judged to rule out the possibility of finding better codes. We should also note that, in our comparison, we have not considered the third generation of quantum repeaters, which rely on quantum error correction for handling both channel loss and gate errors. The requirements of such systems is much more stringent than the ones we considered here, and, at least, in the short term, our above conclusion may be the most relevant one for physical implementation of the system.

Another interesting observation in Fig.~\ref{fig:3dplot}(a) is that the typical values required for $1-F_0$, $\beta$, and $\delta$, in order to offer an advantage over probabilistic repeaters, seems to be quite within a feasible range. For the key error parameter of $\beta$, up to 2\% is acceptable, whereas for fidelity we are looking at lower ninety's, which both seem achievable with current technology. The measurement error can also be kept below 1\%. All in all, our analysis suggests that extending the reach of trust-free terrestrial QKD links to 1000~km is within reach in the near future.

\section{Conclusions}
\label{conclusion_section}
In this work, we benchmarked the performance of a QKD system that relied on quantum error correction for entanglement distillation against probabilistic quantum repeaters that do not necessarily use any additional distillation techniques. In order to improve system performance and simplify its implementation requirements, in the former case, we first proposed two decoding schemes that did not need any two-qubit gates. This reduced the decoding errors, as compared to conventional error-correction decoders, and increased the resilience of the system to common sources of error. In order to analyse the system, we also developed several numerical and analytical approximation techniques, and checked them against exact results in certain cases. This allowed us to study the performance of two codes from the family of repetition codes. We interestingly found that, for most practical purposes, the three-qubit system could offer the best performance so long as error parameters are around 1\%. In higher error regimes, probabilistic quantum repeaters could already offer better rates, or secret key exchange was not at all possible. Our results also shed light into how encoded quantum repeaters would compare with some other classes of quantum repeaters that relied on probabilistic entanglement distillation techniques. We showed that for moderate to high nesting levels the encoded setup could tolerate more errors than the BDCZ protocol. We note that the extension of our decoding and approximation techniques are in principle possible to larger codes, but, in the case of QKD, this may not offer additional advantage. Based on our analysis, it seems feasible to employ current technologies for quantum memories to demonstrate this encoded class of repeaters.

\begin{acknowledgments}
The authors would like to thank Daniel Alsina Leal, Hermann Kampermann, and Dagmar Bru{\ss} for fruitful discussions. This project is funded by the European Union's Horizon 2020 research and innovation programme under the Marie Sklodowska-Curie grant agreement number 675662 (QCALL) and UK EPSRC Grant EP/M013472/1. All data generated in this work can be reproduced by the provided methodology and equations.
\end{acknowledgments}

\appendix

\section{Equivalence of Decoders 2 and 3}
\label{sec:App}
In this Appendix, we prove that decoders 2 and 3, in the ideal case, are equivalent. That would then allow us to use the same security proof that we have for decoder 2, i.e., that of entanglement-based BBM92 protocol \cite{bennett1992quantum}, to decoder 3 as well. In practice then, both decoders 2 and 3 implement the BBM92 protocol with erroneous decoders. Given that these errors result in overestimating bit-flip and phase-flip errors, we can still use the key rate formula in \cref{eq:secfraction} to obtain a lower bound on the secret key rate.

In order to prove our conjecture, we effectively show that the measurement operators implemented, in the ideal case, by either decoders are identical. We first write down the perfect measurement operators for decoder 3 corresponding to measuring bit 0, or, equivalently, states $|0\rangle$ in $Z$ and $|+\rangle$ in $X$ bases, as follows
\begin{align}
    M_0^{(3)} &= |000\rangle\langle 000| + |100\rangle\langle 100| + |010\rangle\langle 010| + |001\rangle\langle 001|,  \label{eq:m0_alt} \\
    M_+^{(3)} &= |+++\rangle\langle +++| + |+--\rangle\langle +--| \nonumber\\
        &+ |-+-\rangle\langle -+-| + |--+\rangle\langle --+|, 
  \label{eq:m+_alt}
\end{align}
respectively. The proof for bit $1$ can similarly be done. In order to show that the effective measurement operators for decoder 2 are equivalent to \cref{eq:m0_alt} and \cref{eq:m+_alt}, we break the circuit of decoder 2 into two parts (see Fig. \ref{fig:dec_structure}(b)): the first step includes two {\sc cnot} operations and the second step contains the corresponding measurements of three qubits and the flip gate on the first qubit only if the outputs of the other two qubits are $|1\rangle$. We look at this process in a backward way and the corresponding projectors right before the second part can be represented as
\begin{align}
    M_0^{\rm mid} &= |000\rangle\langle 000| + |001\rangle\langle 001| + |010\rangle\langle 010| + |111\rangle\langle 111|,  \nonumber\\
    M_+^{\rm mid} &= |+00\rangle\langle +00| + |+01\rangle\langle +01| \nonumber\\
                   &+ |+10\rangle\langle +10| + |+11\rangle\langle +11|, 
\label{eq:m0m+_alt_step2}
\end{align}
which implies the fact that for QKD measurement in $Z$ basis, the outputs of the second and the third qubits will affect the result of the first qubit, while for QKD measurement in $X$ basis, the outputs of the other two qubits does not matter. If we now go back to the input stage of the decoder, where $\text{\sc cnot}_{1 \rightarrow 2}$ and $\text{\sc cnot}_{1 \rightarrow 3}$ are applied subsequently, \cref{eq:m0m+_alt_step2} will be transformed to
\begin{align}
    M_0^{(2)} &= |000\rangle\langle 000| + |001\rangle\langle 001| + |010\rangle\langle 010| + |100\rangle\langle 100|,  \nonumber\\
    M_+^{(2)} &= (|000\rangle + |111\rangle)(\langle 000| + \langle 111|) \nonumber\\
    &+ (|001\rangle + |110\rangle)(\langle 001| + \langle 110|) \nonumber\\
    &+ (|010\rangle + |101\rangle)(\langle 010| + \langle 101|) \nonumber\\
    &+ (|011\rangle + |100\rangle)(\langle 011| + \langle 100|)
\label{eq:m0m+_alt}
\end{align}
Note that to calculate $M_+^{(2)}$, we represent the first qubit in $\{ |0\rangle, |1\rangle \}$ basis before applying the corresponding {\sc cnot} gates. We can see that $M_0^{(2)} = M_0^{(3)}$ now. For $M_+^{(2)}$, after writing all three qubits in $\{ |+\rangle, |-\rangle \}$ basis, we establish that $M_+^{(2)} = M_+^{(3)}$. The derivation steps are straightforward and are left out. This proves our conjecture.

%\bibliographystyle{apsrev}
%\bibliography{ref}
%

\end{document}